\documentclass[%
 reprint,
 superscriptaddress,
 amsmath,amssymb,
 aps,
 pra,
]{revtex4-2}

\usepackage{graphicx}% Include figure files
\usepackage{dcolumn}% Align table columns on decimal point
\usepackage{bm}% bold math
\usepackage{gensymb}
\usepackage{comment}
\usepackage{color}

\begin{document}

\preprint{APS/123-QED}

\title{Measurement of DC and AC electric fields inside an atomic vapor cell with wall-integrated electrodes}

\author{Lu Ma}
\altaffiliation[Present address: ]{Applied Materials, Inc., Santa Clara, CA 95054, USA}
\affiliation{Department of Physics, University of Michigan, Ann Arbor, Michigan, 48109, USA}
\author{Michael A. Viray}
\affiliation{Department of Physics, University of Michigan, Ann Arbor, Michigan, 48109, USA}
\author{David A. Anderson}
\affiliation{Rydberg Technologies Inc, Ann Arbor, Michigan, 48103, USA}
\author{Georg Raithel}
\affiliation{Department of Physics, University of Michigan, Ann Arbor, Michigan, 48109, USA}
\altaffiliation{Present address: Applied Materials, Inc., Santa Clara, CA 95054, USA}

\date{\today}% It is always \today, today,

\begin{abstract}
We present and characterize an atomic vapor cell with silicon ring electrodes directly embedded between borosilicate glass tubes. The cell is assembled with an anodic bonding method and is filled with Rb vapor. The ring electrodes can be externally connectorized for application of electric fields to the inside of the cell. An atom-based, all-optical, laser-spectroscopic field sensing method is employed to measure electric fields in the cell. Here, the Stark effect of electric-field-sensitive rubidium Rydberg atoms is exploited 
to measure DC electric fields in the cell of $\sim 5$~V/cm, with a relative uncertainty of 10\%.  Measurement results are compared with DC field calculations, allowing us to quantify electric-field attenuation due to 
free surface charges inside the cell. We further measure
the propagation of microwave fields into the cell, using Autler-Townes splitting of Rydberg levels as a field probe. Results are obtained for a range of microwave powers and polarization angles relative to the cell's ring electrodes. We compare the results with microwave-field calculations. Applications are discussed.
\end{abstract}

\maketitle

%\tableofcontents

\section{Introduction}

The introduction of DC and AC electric fields into vapor-cell-based quantum devices currently is of considerable interest. For instance, DC Stark-tuning of Rydberg transition frequencies broadens the frequency range of Rydberg-atom-based microwave  field detectors~\cite{sedlacek12, sedlacek13, fan14, holloway14, patentA}. Auxiliary radio-frequency (RF) fields for heterodyne RF field sensing in vapor-cell devices enhance sensitivity~\cite{Jing2020} and allow RF phase sensing~\cite{Simons2019, Anderson2020, patentB}.
Further, a range of quantum devices that require accurate control of in-vacuum electric fields could, potentially, be implemented inside miniaturized glass cells. These include Paul~\cite{Leibfried2003, ito17}, Penning~\cite{Brown1986, Hall1996, mcmahon20} and cusp~\cite{Enomoto2010} traps for ions, electrons and plasmas, and Faraday-shielded magneto-optical-traps (MOTs)~\cite{beloy2018, nichols20} for applications requiring controlled electromagnetic boundary conditions or a well-defined black-body radiation environment~\cite{ushijima2015,xu2016,golovizin2019}.

DC and low-frequency electric-field control inside glass vacuum cells using electrodes placed outside of the cells is challenging due to cell-internal photo-electric charging effects~\cite{PhysRevLett.98.113003}. While such effects 
can be controlled in order to enable all-optical generation of cell-internal electric fields~\cite{ma20}, physical electrodes inside the cell provide improved electric-field control with greater flexibility in field geometry. Here, we seek structures that are directly integrated into the glass cell wall, simultaneously serving as electrodes and as through-connects.

\begin{figure}
    \centering
    \includegraphics[width=\linewidth]{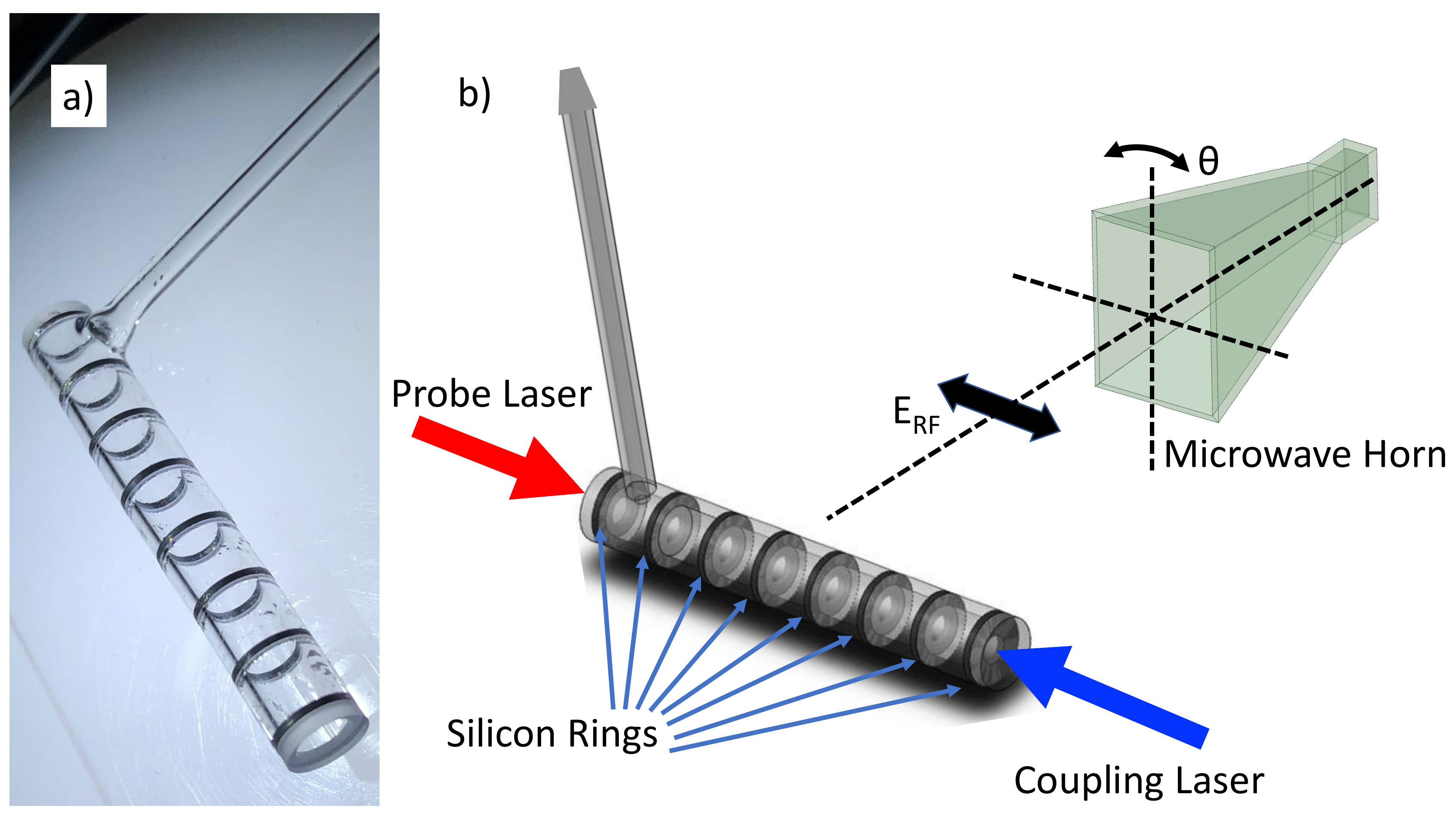}
    \caption{Images of glass vapor cell with wall-integrated electrodes for application of DC and AC electric fields. (a) Picture of the actual cell. The black rings are anodically-bonded conducting-Si rings. (b) Illustration of the testing setup, showing the cell, the rubidium $5S_{1/2} \rightarrow 5P_{3/2}$ EIT probe laser (780~nm, red), the $5P_{3/2} \rightarrow \, Rydberg$  EIT coupler laser (480~nm, blue), and the microwave-emitting horn. The lead wires to the Si rings are only connected for DC field measurements. The setup allows variation of the microwave polarization angle $\theta$. The displayed case is for maximum  microwave transmission into the cell.}
    \label{fig:setup_fig}
\end{figure}

To incorporate electrodes directly into the body of a glass vacuum cell, the electrodes must be attached to the glass in a way that yields a seal between the two materials with sufficiently low leak rates and outgassing levels for use in high vacuum. This places limits on both the electrode material and the possible bonding processes. 
We have found that the integration of highly conductive doped-Si parts into borosilicate glass walls via an anodic-bonding technique~\cite{liew04, kitching18, nishino21, sebbag21} enables the fabrication of high- and ultra-high-vacuum devices. The Si electrodes afford electric-field control inside the cell directly through the cell wall. The electric fields can have complex geometry, and the cells can be integrated onto electronic circuit boards. One cell structure we have fabricated is shown in Fig.~\ref{fig:setup_fig}(a). The cell is mostly made of borosilicate glass (pyrex) and features eight evenly-spaced ring electrodes along the length of the body. Application of different voltage combinations to the eight electrodes allows the realization of a variety of electric-field geometries. The rings also serve a secondary, serendipitous use as a partial polarizer for microwave radiation, thanks to the small 6~mm spacing between them.

We utilize laser spectroscopy of an atomic vapor to analyze the electric fields in the cell.
Atom-based electric field sensing is currently an active field of research, drawing interest from both researchers studying fundamental science, and engineers working on integrated quantum metrology systems. This technique uses atoms in Rydberg states (states with high principal quantum number $n$) as electric ($E$)-field probes. Though electromagnetically-induced transparency (EIT)~\cite{scully, fleischhauer}, the atoms yield  narrow laser-spectroscopic lines that mark the optical excitation energy of the Rydberg levels. Rydberg-level Stark shifts can be observed
via EIT line shifts. From measured line shifts and splittings one can determine the strength of the $E$-field in the laser-probe region. This technique, which takes advantage of the large susceptibilities of Rydberg atoms to DC and AC electric fields~\cite{Dunning, Gallagher}, has previously been used to observe a variety of electric fields, including radio HF and VHF fields~\cite{bason10,miller16,Jiao2016}, monochromatic~\cite{sedlacek12, sedlacek13, fan14, holloway14} or modulated~\cite{receiver} microwave radiation, plasma electric fields~\cite{anderson17, weller19}, and DC electric fields~\cite{bason10, Grimmel2015, ma20}.

We first study a DC electric-field configuration by applying a variable DC voltage to one of the Si ring electrodes, with the others grounded. In the second configuration, a microwave field is injected into the cell by radiating microwaves of variable intensity and polarization from a microwave horn into the cell. In each configuration, we analyze Rydberg-EIT signals to deduce the electric field. 

\section{Experimental Setup}

\subsection{Cell Fabrication}

In the late 1960s, G. Wallis and D. I.Pomerantz (P. R. Mallory \& Co. Inc.)
demonstrated a glass-metal sealing method with the assistance of a static electric field~\cite{1970110112284}, which became known as anodic bonding ~\cite{madou2018fundamentals,1185603,2000295196843}.
We employ anodic bonding of borosilicate glass tubes and highly-doped Si rings to create buffer-gas-free Rb vapor cells with wall-integrated electrodes. Glass surfaces are prepared with a standard Chemical-Mechanical-Polishing (CMP) process. The Si rings are fabricated from double-side polished Si wafers. The bonding process is performed at a temperature of about 300$^\circ$C. The bonding voltage applied to the Si rings ranges from 600 to 900~V, limited
by sparking at higher voltages.

\begin{figure}
    \centering
    \includegraphics[width=\linewidth]{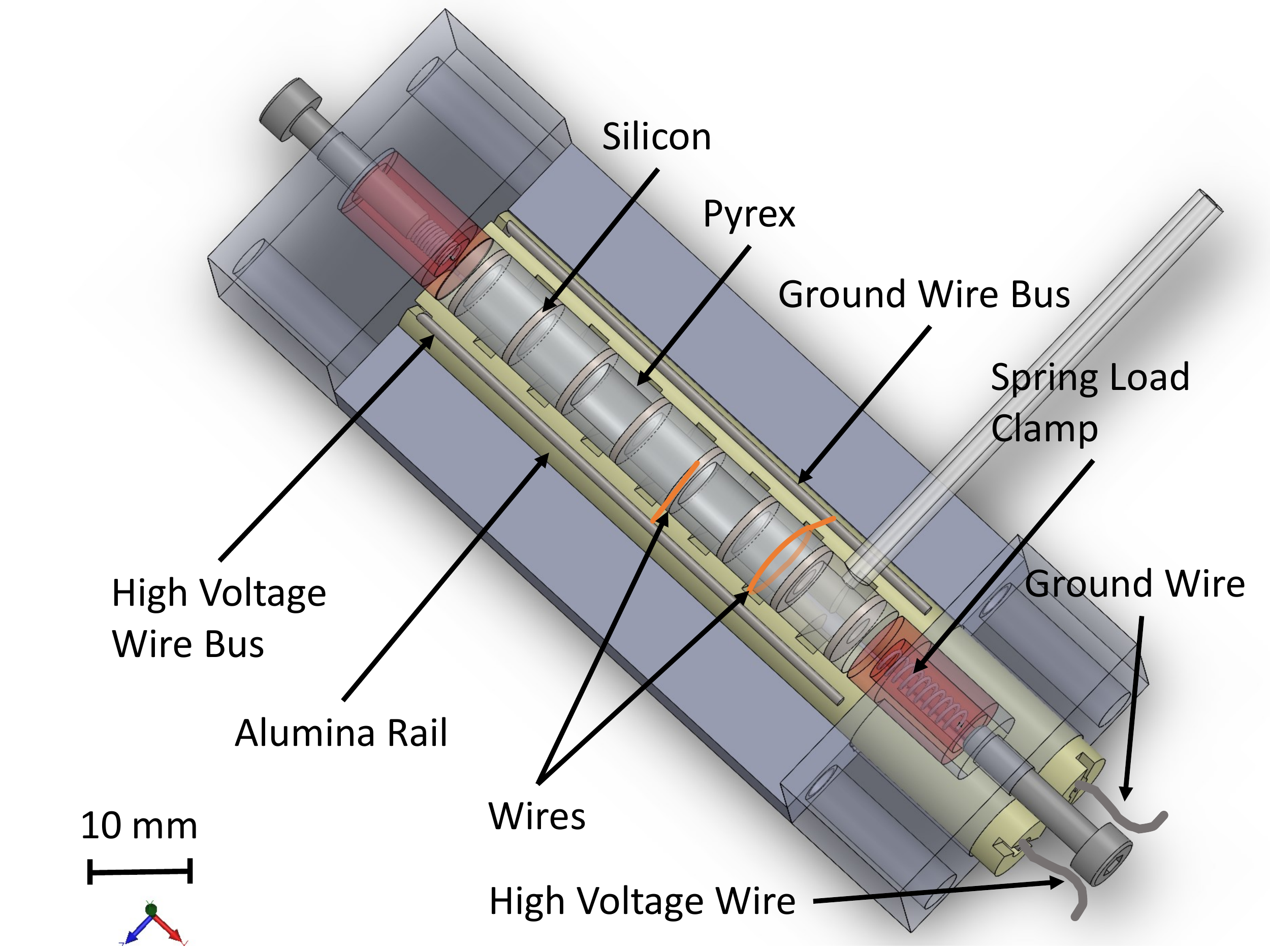}
    \caption{Bonding apparatus used for simultaneous anodic bonding of 9 borosilicate glass parts with 8 highly-conductive Si rings. Bonding voltages are applied through thin wires connectorized with solid-wire buses.}
    \label{fig:bonding_jig}
\end{figure}

In order to minimize the number of thermal cycles, our multi-surface anodic bonding is performed in one step using a custom-designed, compact bonding jig shown in Fig.~\ref{fig:bonding_jig}.
The circular bonding pieces, which all have equal diameters, are assembled and aligned along a common axis on top of two supporting ceramic tubes. To ensure contact across the 16 bonding surfaces, the pieces are clamped between spring-loaded caps with setscrews for force adjustment. The Si rings and glass parts are wrapped with thin wires, which are connected to metal rails that pass through the centers of the slotted ceramic tubes. The bonding voltages are applied via the rails. During the bonding, the jig is heated on a hot plate and covered with an Al sheet-metal lid. After bonding, the cell is filled with Rb vapor at high-vacuum pressure and sealed.  

\subsection{Optical and Electric-Field Setup}

Figure~\ref{fig:setup_fig}b shows the optical and microwave / DC electric-field setup. We use a two-photon EIT ladder configuration  in which the uppermost state is a Rydberg level. The two counter-propagating EIT lasers are overlapped along the cell axis,  as illustrated in Fig.~\ref{fig:setup_fig}~(b). The first laser drives the $5S_{1/2} \rightarrow 5P_{3/2}$ EIT probe transition and has a wavelength of $\lambda_p=780$~nm (red), a power of 11~$\mu$W, and a diameter of $w_0 = 464$~$\mu$m. The second laser (EIT coupler), runs at $\lambda_c=480$~nm (blue) and 46~mW, has a diameter of $w_0 = 268$~$\mu$m,  and drives the $5P_{3/2} \rightarrow 49D_{5/2}$ (AC field measurements) or the $5P_{3/2} \rightarrow 32S_{1/2}$ (DC field measurements) Rydberg transitions. The EIT signal is obtained by scanning the coupler laser across the Rydberg resonances. The probe beam is sent into a photodiode connected to a transimpedance amplifier. The EIT signal, displayed on and acquired with an oscilloscope, exhibits Rydberg transitions whose quadratic DC and linear AC Stark effects reveal the electric fields present in the cell.  

The DC electric fields are applied by mounting the cell on a small piece of circuit board and connectorizing the Si rings to external voltage-control lines, as shown in some detail in Figs.~\ref{Fig:DC_Test_Setup}(a) and~(b).

In the microwave measurements, the vapor cell is mounted stem-down on a thin anti-static acetal rod, which is held in place by a post holder. The Si rings are floating, with no wires attached. The microwave radiation is emitted from a 15-dBi standard gain microwave horn (Pasternack~PE9852/2F-15) connected to a synthesizer. We set the microwave frequency to 18.149805~GHz to resonantly drive the $49D_{5/2} \rightarrow 50P_{3/2}$ Rydberg transition. The horn is attached to a rotating mount to allow for a variation of the microwave polarization angle $\theta$ (see Fig.~1~(b)) over a range of 360$^\circ$, with a resolution of 1$^\circ$. The microwaves are incident on the side of the vapor cell, which is in the horn's far field, at a distance of 15~cm.

\section{Experimental Analysis}
\label{sec:experimental}

\subsection{DC electric field}

\begin{figure}
    \centering
    \includegraphics[width=\linewidth]{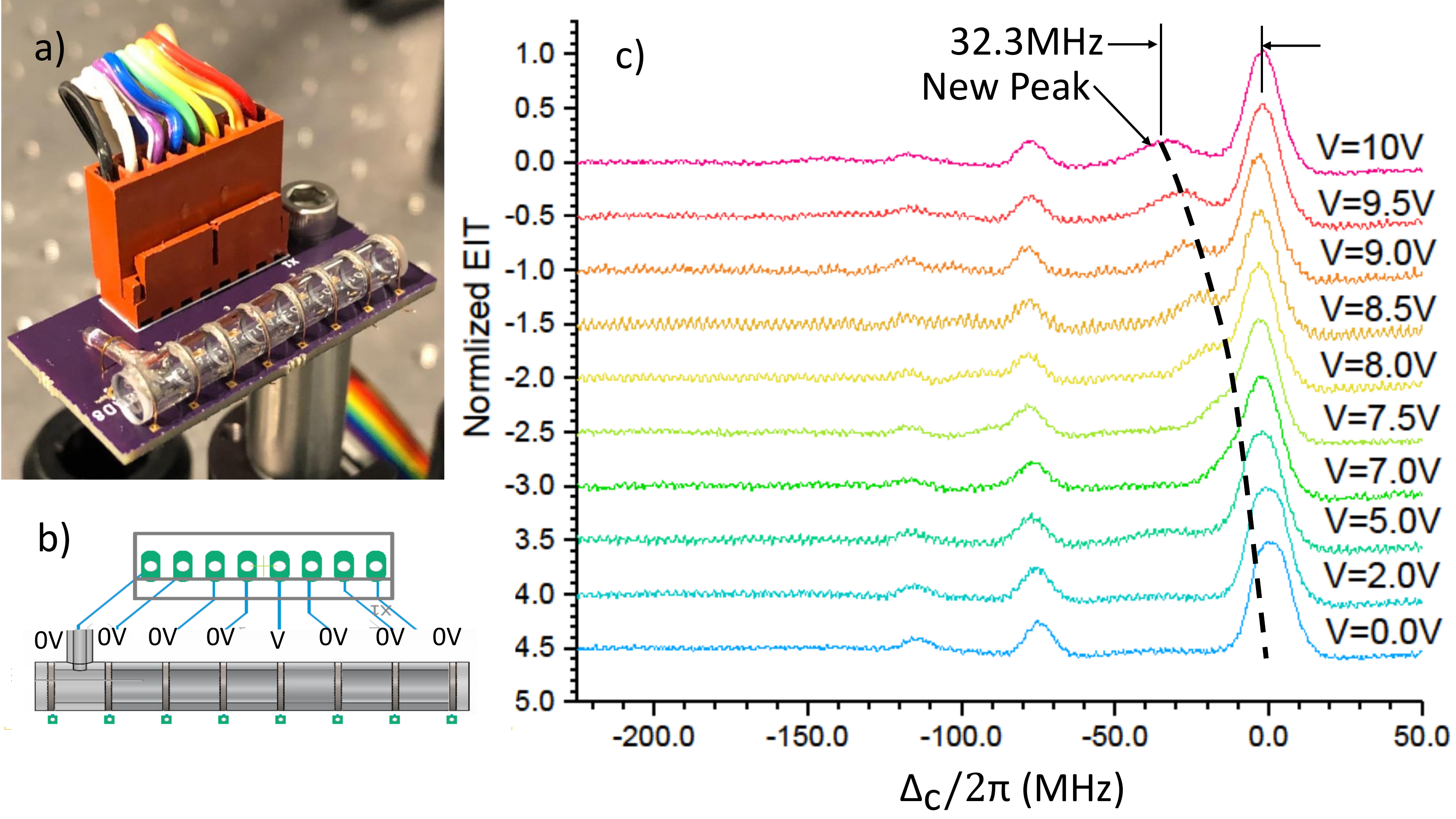}
    \caption{(a) Picture of the cell setup used for spectroscopic  DC electric-field measurement in the cell. Individual voltages are applied to the highly-conductive Si ring electrodes through the ribbon cable  and thin wires glued} to the outer rims of the Si rings with conductive glue. (b) Wiring diagram used for the data in (c). (c) Rydberg-EIT spectra of Rb $31S_{1/2}$ for the indicated voltages $V$ vs detuning of the coupler laser from the $5P_{3/2}$, $F=4$ $\rightarrow$ 
    $31S_{1/2}$ transition. The ``new peak'', most clearly visible on the main line near $0$~MHz, is due to DC Stark shifts near the voltage-carrying Si ring.
    \label{Fig:DC_Test_Setup}
\end{figure}

First, we test our device for DC electric field injection capability using the cell setup shown in Fig.~\ref{Fig:DC_Test_Setup}(a). The strength of the DC field inside the cell is monitored through the DC Stark shift of the Rydberg EIT signal. In this test, the 32S$_{1/2}$ Rydberg state, which has a DC polarizability $\alpha$ of 2.214 MHz/(V/cm)$^2$, is used. The shift in a DC field $E_{DC}$ is given by $-\alpha E_{DC}^2/2$.

When the voltage $V$ applied to the center Si ring is increased (Fig.~\ref{Fig:DC_Test_Setup}(b)), a new EIT peak starts to emerge on the red side of the strongest EIT resonance
(Fig.~\ref{Fig:DC_Test_Setup}(c)). This new peak arises
from DC Rydberg level shifts in the region around the center Si ring, where the applied voltage $V$ induces DC electric fields. The maximum frequency separation between the shifted peak (``New peak'' in Fig.~\ref{Fig:DC_Test_Setup}(c)) 
and the main peak is ($32.3\pm5$)~MHz, corresponding to a measured DC electric-field maximum of $E_{DC} = (5.4 \pm 0.5)$~V/cm. We note that in Fig.~\ref{Fig:DC_Test_Setup}(c) there are several copies of the same Rydberg spectrum near $0$~MHz, $-75$~MHz and $-115$~MHz  that belong to the $F=4$, $F=3$ and $F=2$ hyperfine states of the intermediate $5P_{3/2}$ level, respectively. The hyperfine splittings are reduced by a factor $\lambda_p/\lambda_c -1$ due to EIT Doppler shifts in the cell. 

\begin{figure*}
    \centering
    \includegraphics[width=0.8\textwidth]{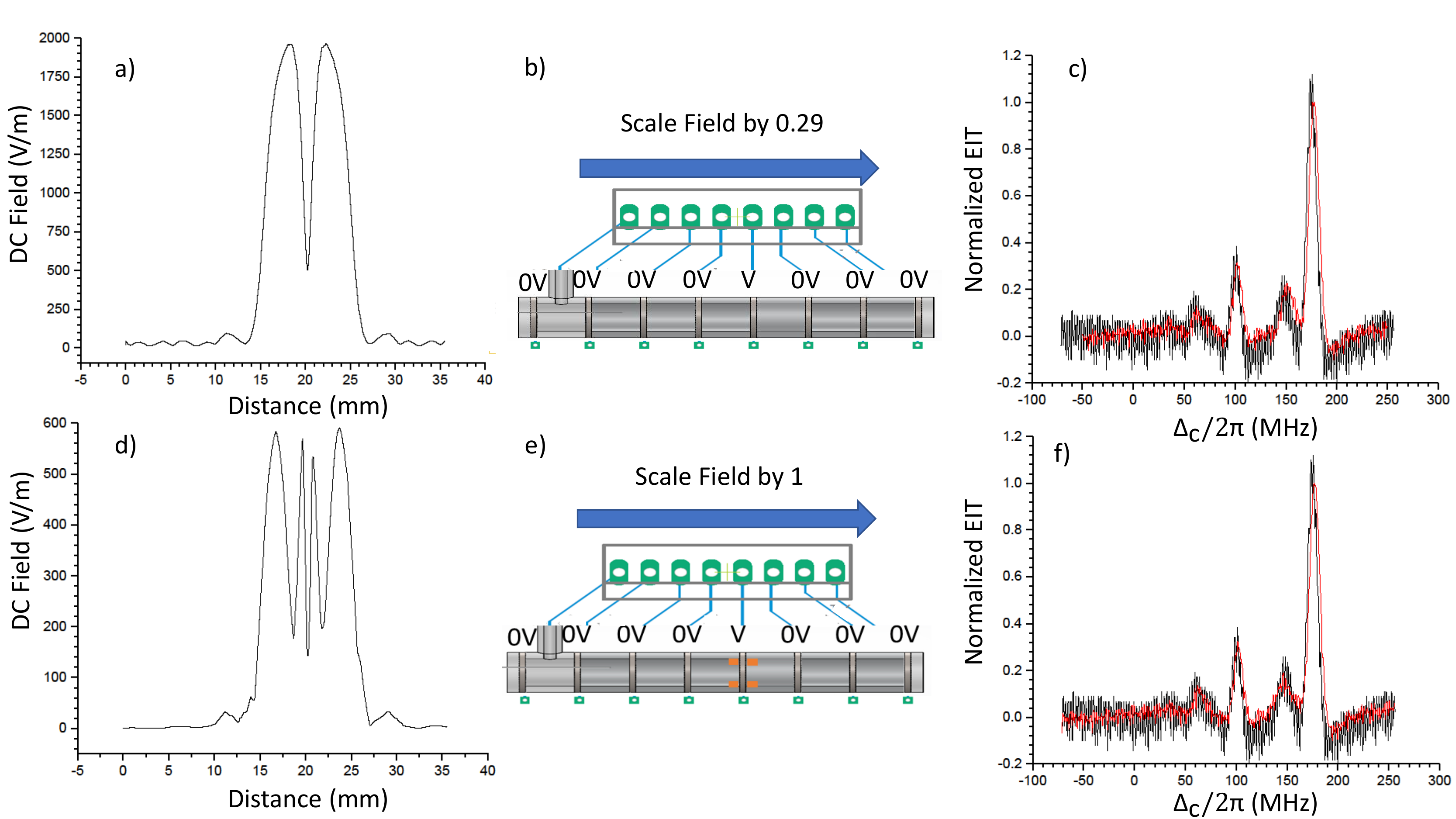}
    \caption{Calculated DC electric fields for an unshielded (a) and a shielded (d) case for $V=10$~V. Simulated DC Stark spectra are plotted in red in (c) and (f), respectively, measured spectra are plotted in black. The calculated unshielded DC field from (a) must be scaled down by a factor of $\approx 3$, as visualized in (b), to arrive at a reasonable match between simulated and measured spectra in (c). In a simulation of a physical field-reducing mechanism, we add two grounded metal sheaths (orange parts in (e)) on both sides of the voltage-carrying Si ring. This leads to reasonable agreement 
    between simulated and measured spectra (f), without any further field attenuation.}
    \label{Fig:DC_Sim_Plt}
\end{figure*}

In Fig.~\ref{Fig:DC_Sim_Plt} we show calculated distributions of the DC electric field, $E_{DC}$, along the cell axis and compare modeled with experimentally measured Stark-shifted EIT spectra. In an initial model, we compute the electric-field distribution in the cell without any surface charges on the class walls (Fig.~\ref{Fig:DC_Sim_Plt}~(a)). We then compute an average of Stark-shifted EIT spectra of the $33S_{1/2}$-level, with the average taken over the computed distribution of $E_{DC}$ along the cell axis. The initial model predicts a quadratic Stark shift that is about a factor of ten larger than measured. A global attenuation of $E_{DC}$ by a factor of $\approx 3$, as visualized in Fig.~\ref{Fig:DC_Sim_Plt}~(b), results in a simulated spectrum  (red line in Fig.~\ref{Fig:DC_Sim_Plt}~(c)) that closely resembles the measured one (black line). In light of this finding, we then calculate the $E_{DC}$-distribution for a 
physical model in which the voltage-carrying Si ring is straddled by two grounded rings (orange sheaths in  Fig.~\ref{Fig:DC_Sim_Plt}~(e)) on the inner cell wall. The grounded rings simulate the field-reducing effect of surface charges accumulated on the inner glass surfaces near the voltage-carrying Si ring. The resultant simulated spectrum in Fig.~\ref{Fig:DC_Sim_Plt}~(f)  (red line) closely approaches the measured spectrum (black line), without any additional field attenuation needed for agreement. 

Future spatially-resolved Rydberg-EIT DC electric-field measurements may reveal more detail as to how exactly the field reduction by a factor of $\approx 3$ occurs.
The electrode structure as well as the overall cell symmetry suggest that $E_{DC}$ should have a leading quadrupole term, as is also evident in the field simulations. The ability to create such fields, using one or more of the wall-integrated electrodes, will be valuable for confining charged particles in vapor-cell Penning or cusp traps, which would require an additional longitudinal magnetic field, or Paul traps, which would utilize voltage drives in the HF and VHF radio-frequency regimes.

\subsection{Injected microwave electric fields}
\label{subsec:experimental_mw}

\subsubsection{Varied Microwave Power at Fixed Polarization Angle}
\label{subsubsec:varying_power}

\begin{figure}
    \centering
    \includegraphics[width=0.7\linewidth]{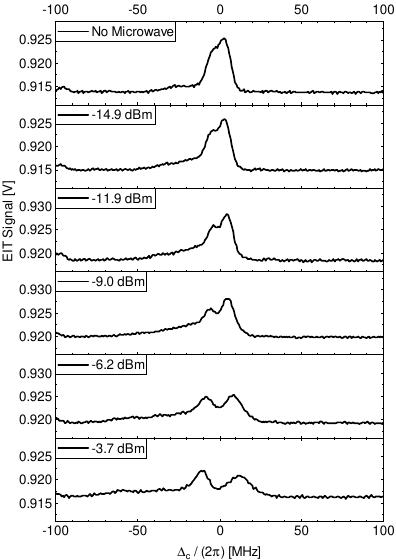}
    \caption{EIT signals vs coupler laser detuning for the indicated microwave powers. For these data, the microwave polarization angle is set for maximum transmission ($\theta = 0$).}
    \label{fig:powerstack}
\end{figure}

In a first set of microwave measurements, the horn is kept at $\theta = 0^\circ$ (see Fig.~1~(b)), the microwave polarization angle of maximum transmission. The Si rings are on floating DC potentials and are not connected to any wires. We measure resonant Autler-Townes (AT) splittings of the $49D_{5/2} \rightarrow 50P_{3/2}$ Rydberg transition at 18.149805~GHz, for microwave powers injected into the horn of $P_{horn}=$-3.7, -6.2, -9.0, -11.9, and -14.9~dBm, as measured by a power meter. Additionally, we record a control measurement where the microwave is turned off. In Fig.~\ref{fig:powerstack} we show the EIT spectra recorded at these values of $P_{horn}$. Since this measurement qualitatively differs from earlier work due to the presence of the Si rings, in the following we analyze the data in some detail.

\begin{figure}
    \centering
    \includegraphics[width=0.9\linewidth]{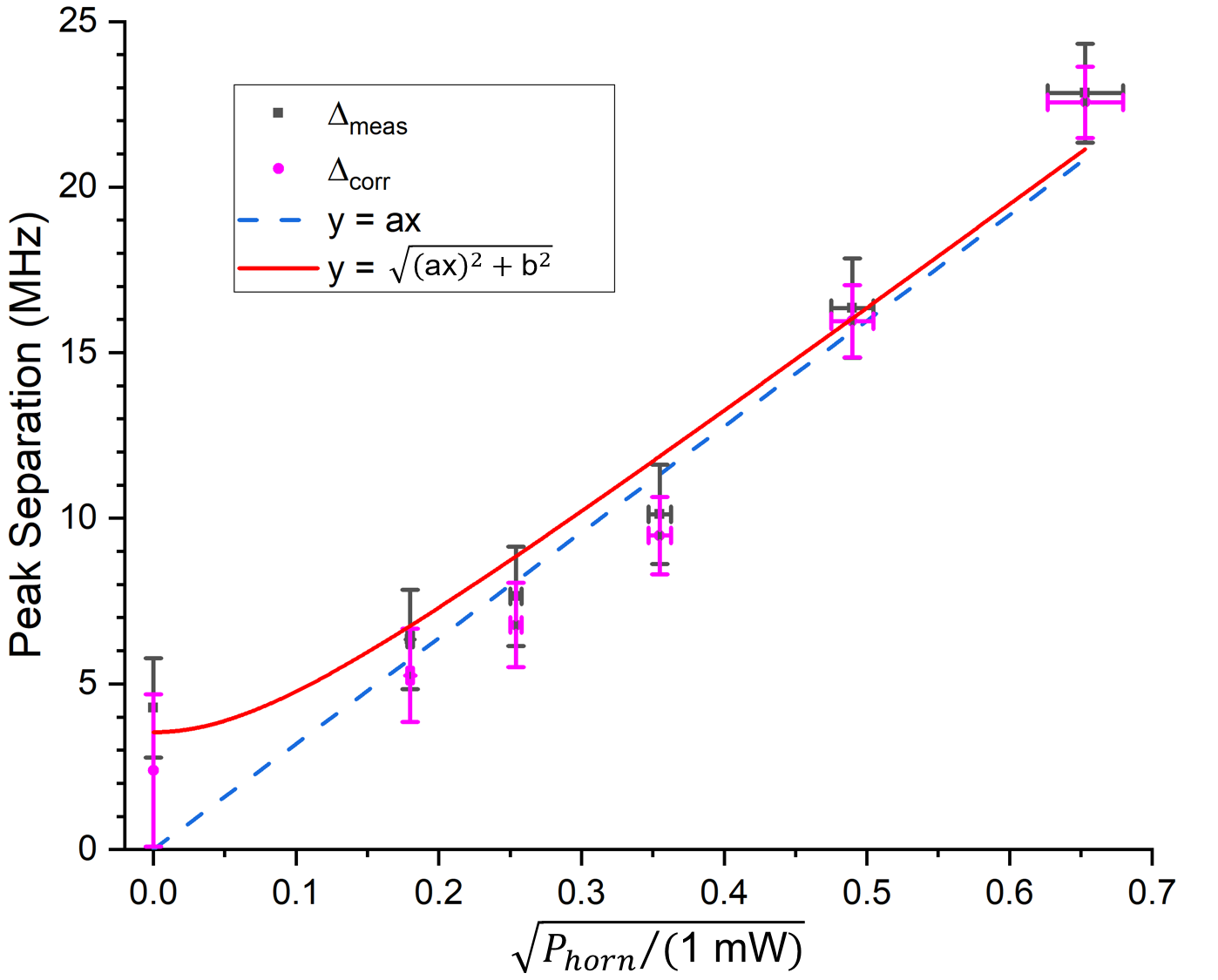}
    \caption{Autler-Townes splitting of the EIT signal in MHz vs $x=\sqrt{P_{horn}/1~{\rm {mW}}}$. The black data points are measured peak separations, $\Delta_{meas}$. The solid red curve is a fit for $\Delta_{meas}$ of the form $y = \sqrt{(ax)^2 + b^2}$ with fit parameters $a$ and $b$ (see Eq.~\ref{align:parameters}). The dashed blue line shows $y = ax$. The magenta data points are corrected peak separations, $\Delta_{corr}$ (see Eq.~\ref{eq:corrected}).}
    \label{fig:splittingvsfield}
\end{figure}

In Fig.~\ref{fig:splittingvsfield} we plot the AT splitting in MHz as a function of $\sqrt{P_{horn}/1~{\rm {mW}}}$. The black data points are the measured EIT line splittings, denoted $\Delta_{meas}$. The splittings are due to several effects. The microwave electric field, $E_{MW}$, induces an AT splitting that ideally is proportional to $E_{MW}$ and that represents the dominant contribution to $\Delta_{meas}$ at the higher powers in Fig.~\ref{fig:splittingvsfield}. Further, Fig.~\ref{fig:powerstack} shows that there is a residual splitting of several MHz, denoted $b$, even when the microwave is turned off. The residual splitting is indicative of DC stray electric fields from the cell walls, and possibly a Zeeman splitting due to stray magnetic fields. We empirically quantify the splitting effects by applying a fit to the black data points of the form $\Delta_{meas} = \sqrt{(ax)^2 + b^2}$, where $x=\sqrt{P_{horn}/{\rm mW}}$. The term $a x$ accounts for the microwave-induced ideal AT splitting of the Rydberg line, which is dominant at higher microwave powers $P_{horn}$. The $b$-term accounts for the stray-field-induced splitting, which is dominant at zero and low values of $P_{horn}$. 

The fit result is shown in Fig.~\ref{fig:splittingvsfield} in solid red. 
The $a$ and $b$ values from the fit are:
\begin{align}\label{align:parameters}
    a &= (31.9 \pm 1.7) \textrm{ MHz} / \sqrt{P_{horn}/{\rm mW} } \nonumber \\
    b &= (3.54 \pm 0.90) \textrm{ MHz}
\end{align}
The dashed blue line in Fig.~\ref{fig:splittingvsfield} shows the ideal microwave-induced AT splitting, $a x$, that the Rydberg line would exhibit in a Doppler-free cold-atom measurement~\cite{holloway2017} with no DC fields present. The magenta data points are corrected line splittings, $\Delta_{corr}$, in which we subtract off the splitting contribution due to the DC stray field according to:
\begin{equation}\label{eq:corrected}
    \Delta_{corr} = \sqrt{\Delta_{meas}^2 - b^2} \approx a x
\end{equation}
The value of $\Delta_{corr}$ has an improved linear relationship with the microwave field $E_{MW}$.

The corrected splitting, $\Delta_{corr}$, approximately equals the Rabi frequency, $\Omega_0$, of the resonantly driven microwave transition (in Hz). The following equation applies,
\begin{equation}
\label{eq:emw}
    E_{MW} d / h = \Omega_0 = a x \approx \Delta_{corr} \quad .
\end{equation}
Here, $d$ is the transition dipole moment, and $h$ is Planck's constant. %If $d$ and $\Omega_0$ are known, then the electric field for each splitting can be calculated. 
Our applied microwave field drives the $49D_{5/2} \rightarrow 50P_{3/2}$ Rydberg transition, which has a known transition dipole moment of $d = 1.16 \times 10^{-26}$~Cm. Equation~\ref{eq:emw} and the corrected splittings $\Delta_{corr}$ then allow us to calculate the electric field $E_{MW}$.  

\begin{table}
    \centering
    \begin{tabular}{c|c|c|c}
         $\sqrt{P_{Horn}/1\textrm{mW}}$ & $\Delta_{meas}$ [MHz] & $E_{MW}$ [V/m] & $x / E_{MW}$ [m/V] \\
         \hline
%         0.158 & 4.27 & 0.243 & 0.0370 \\
         0.18 $\pm$ 0.01 & 6.3 $\pm$ 1.5 & 0.36 $\pm$ 0.09 & 0.50 $\pm$ 0.12\\
         0.25 $\pm$ 0.01 & 7.7 $\pm$ 1.5 & 0.44 $\pm$ 0.09 & 0.58 $\pm$ 0.12\\
         0.36 $\pm$ 0.01 & 10 $\pm$ 1.5 & 0.58 $\pm$ 0.09 & 0.62 $\pm$ 0.09\\
         0.49 $\pm$ 0.02 & 16 $\pm$ 1.5 & 0.93 $\pm$ 0.09 & 0.53 $\pm$ 0.05\\
         0.65 $\pm$ 0.02 & 23 $\pm$ 1.5 & 1.3 $\pm$ 0.09 & 0.50 $\pm$ 0.04
    \end{tabular}
    \caption{The columns show 
$x=\sqrt{P_{Horn}/1\textrm{mW}}$, the measured splittings $\Delta_{meas}$, the electric fields $E_{MW}$ calculated from Eqs.~\ref{eq:corrected} and~\ref{eq:emw}, and the ratio $x / E_{MW}$, which should ideally be constant.}
    \label{tab:varying_power}
\end{table}

A summary of results is shown in Table I. Overall, the data series shows that $E_{MW}$ can be measured in our cell with wall-integrated Si electrodes using established methods for atom-based microwave electric-field sensing~\cite{sedlacek12, sedlacek13, fan14, holloway14, holloway2017}. The field injected into the cell is only about $60\%$ of the field that would be present without the cell (see details in Sec.~\ref{sec:simulation}). In view of anticipated electron-, ion- and plasma-trapping applications of this cell, it mostly is important to conclude that microwave field injection into the cell for plasma diagnostics and charged-particle-drive will be efficient. A detailed evaluation of the 
cell for the purpose of microwave field metrology is not part of the present work.

\subsubsection{Varied Polarization Angle at Fixed Microwave Power}
\label{subsubsec:varying_angle}

\begin{figure}
    \centering
    \includegraphics[width=0.7\linewidth]{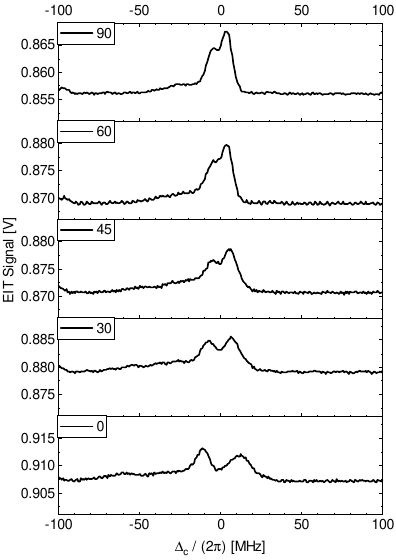}
    \caption{EIT signals vs coupler-laser detuning for the 
    indicated values $\theta$ of the incident microwave polarization. The power is fixed at $P_{horn}=-3.7$~dBm, the highest value in Fig.~5.}
    \label{fig:anglestack}
\end{figure}

\begin{figure}
    \centering
    \includegraphics[width=0.9\linewidth]{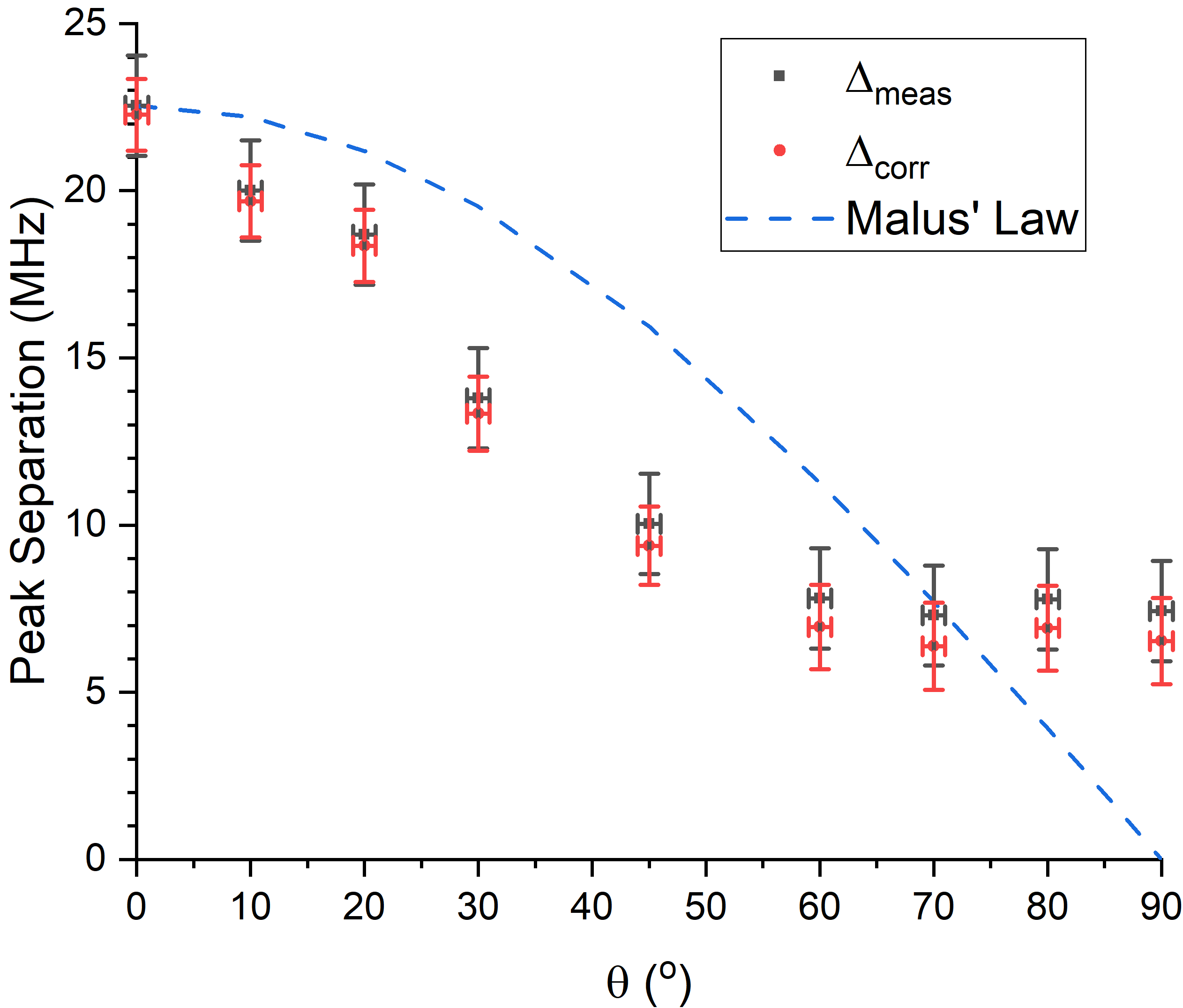}
    \caption{Splittings $\Delta_{meas}$ (black) and  $\Delta_{corr}$ (pink) vs polarization angle of the incident microwave radiation, with the relation between  $\Delta_{meas}$ and  $\Delta_{corr}$ established in Eq.~\ref{eq:corrected}. For reference, the dashed blue curve represents Malus's law, as it would apply for an ideal microwave polarizer.}
    \label{fig:splittingvsangle}
\end{figure}

In our second set of microwave measurements we use a fixed power
$P_{horn}=-3.7$~dBm and the horn is rotated. 
We take measurements of the AT splitting from $\theta=0^\circ$ (maximum transmission) to $\theta=90^\circ$ (minimum transmission), going in increments of about 10$^\circ$. In Fig.~\ref{fig:anglestack} we show the AT splitting of the EIT signal at selected angles, and in Fig.~\ref{fig:splittingvsangle} the measured (black) and corrected (red) splittings in MHz for all angles. We see that, as expected, the splitting decreases as the polarization of the microwave-field direction becomes aligned parallel with the conducting rings.

It is instructive to compare the observed polarization dependence with that of an ideal microwave polarizer (such as one formed by a large, planar grid of parallel wires with a wire separation much less than the microwave wavelength). The transmission behavior of the ideal microwave 
polarizer follows Malus's law~\cite{halliday}, {\sl{i. e.}} the transmitted electric field, $E_{MW}$, would be
\begin{equation}\label{eq:polarization}
    E_{MW} = E_0 \cos(\theta) \quad.
\end{equation}
There, $E_0$ is the field without polarizer, and the angle $\theta$ is defined such that maximum transmission occurs at $\theta=0^\circ$. For comparison, we have added a blue, dashed curve in Fig.~\ref{fig:splittingvsangle} that represents the behavior of an ideal microwave polarizer. 

Comparing the measured and corrected separations, $\Delta_{meas}$ and $\Delta_{corr}$, with the Malus's-law curve, we see that the separations not only decrease faster with $\theta$ for $\theta \lesssim 30^\circ$, but they also level out at about 25$\%$ of the $\theta=0$-splitting when $\theta$ approaches 90$^\circ$. Evidently, the cell electrodes do not act as a perfect microwave polarizer, as explained more in the following paragraph. In view of applications that require external microwave injection, the results show that microwave fields polarized transversely relative to the cell axis still transmit at about 25$\%$. This is relevant, for instance, if one intended to drive the electron cyclotron motion in a vapor-cell Penning trap formed with the cell. This trap would have a strong magnetic field along the cell axis, and the cyclotron motion would be transverse, requiring a transversely polarized ($\theta =90^\circ$) microwave field to drive it.   

There are several reasons for why the cell is not an ideal microwave polarizer. Mostly, the cell structure - finite conductive-Si rings embedded in a dielectric with large dielectric constant - exhibits a microwave response that
differs substantially from that of a grid of parallel wires in vacuum. This is shown in the microwave field simulations below. The simulations also bear out that the field polarization inside the cell may have some ellipticity. A minor effect that may contribute to the measurement result is that, while the microwave horn is perfectly polarized at our level of precision, reflections of microwaves from parts of the setup may be polarized differently from the incident wave and may enter into the cell.
We can discard another potential cause for non-ideal polarizer behavior, namely that the skin depth in the Si could, in principle, be too large. For the wafer grade we use (resistivity 0.001 to 0.005~$\Omega$cm), we estimate a skin depth $\lesssim 10~\mu$m at 18~GHz, which is 1/50 of the wafer thickness. Hence, the Si rings behave as near-ideal conductors, as confirmed in our microwave simulations.

\section{Microwave Simulations}
\label{sec:simulation}

To gain more insight into the microwave experiment discussed 
in Sec.~\ref{subsec:experimental_mw}, we simulate the horn plus cell system using Ansys HFSS software for the geometric parameters used in the experiment. The model further uses accurate dielectric constants and resistivities. Here we show results for $\theta=0^\circ$ and $90^\circ$. We also run simulations without the cell present to calibrate the in-cell microwave fields relative to the cell-free case. 

\begin{figure*}
    \centering
    \includegraphics[width=13 cm]{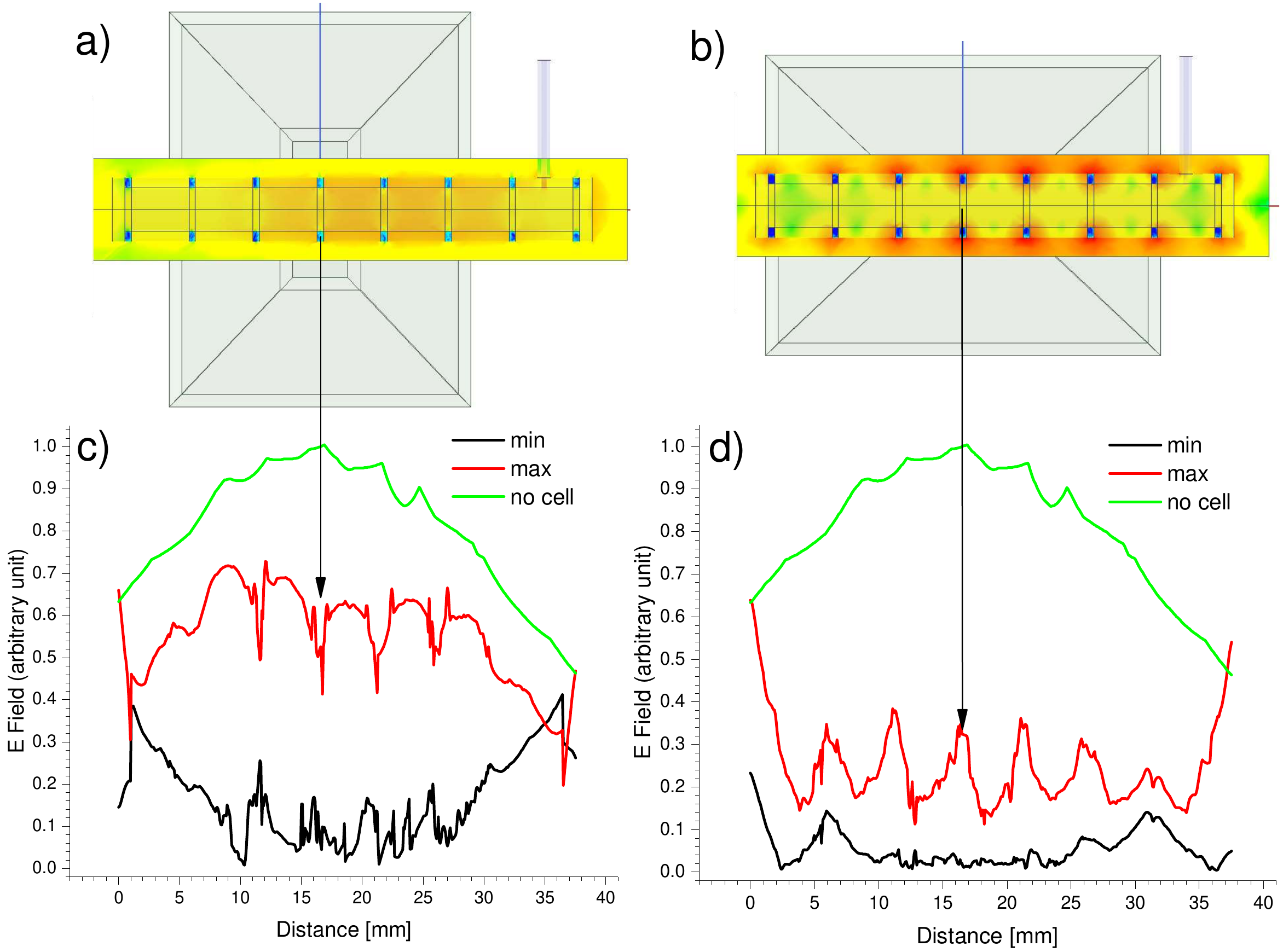}
    \caption{Results of an Ansys HFSS simulation of the microwave experiment described in Subsec.~\ref{subsec:experimental_mw}. a) and b) show horn and cell geometries, as well as the calculated electric-field distributions on planes containing the cell axis (colors explained in text). c) and d) show the field-strength magnitudes on the cell axis at temporal phases of maximal (red) and minimal (black) field strength, and for the case that the cell is absent (green).}
    \label{fig:RF_Sim}
\end{figure*}

Figs.~\ref{fig:RF_Sim}(a) and (b) show the geometries for $\theta=0^\circ$ and $90^\circ$. The coloring inside the cell indicates the microwave electric-field magnitude at the temporal phase when the field is greatest, on a color scale ranging from green (weak) to red (strong field).  
Fig.~\ref{fig:RF_Sim}(c) and (d) show the field magnitude along the cell axis at the temporal phases of maximal (red) and minimal (black) fields. The green curve shows the maximal field without cell present. (The field minimum without cell present is practically zero, because the incident field is linearly polarized).

Since the experimental data represent an average of the microwave field along the length of the cell, we look at the behavior of the simulated-field averages.
Fig.~\ref{fig:RF_Sim}(a) and (c) ($\theta = 0$, corresponding to largest transmission overall),
show that the electric-field transmission, averaged along the cell's axis, is about $60\%$ of what it would be if no cell were present. Fig.~\ref{fig:RF_Sim}(b) and (d) ($\theta = 90^\circ$, corresponding to lowest transmission overall), indicate a transmission of about $25\%$. The simulated average-field ratio been the two cases therefore is about 2.5:1, whereas in the experiment it is about 3:1. We therefore see good qualitative agreement between the simulated and measured field ratios. Deviations are mostly attributed to fabrication uncertainties in the cell geometry.  

While the experiment in its present form only yields a field average, the simulation has sub-mm spatial resolution. The electric field distributions in Figs.~\ref{fig:RF_Sim}(a) and (c) show that for $\theta = 0^\circ$ (high transmission) the field tends to be larger in the spaces between the rings and weaker in the planes of the rings. For $\theta = 90^\circ$ (low transmission; Figs.~\ref{fig:RF_Sim}(b) and (d)), the opposite trend is observed. Another noteworthy detail is that the field minima (black curves) are substantial ($5\%$ to $10\%$ of the cell-free field, in both cases of $\theta$). This is indicative of a slightly elliptical character of the field's polarization inside the cell. The ellipticity is attributed to the fact that the cell components are neither infinitely long parallel wires, nor is the space between the rings filled by vacuum (as it would be for a parallel-wire microwave field polarizer). It is therefore plausible to assume that the structural parts scatter the field into directions orthogonal to the incident field, with material-related phase shifts. This can lead to elliptically polarized net fields on the cell axis, as suggested by the black and red curves in Figs.~\ref{fig:RF_Sim}(c) and~(d).

\section{Conclusion}

Using an anodic-bonding method, we have fabricated a vapor cell that incorporates conductive Si electrodes directly into the body of a high-vacuum Rb vapor cell. DC electric fields applied via the electrodes and microwave electric fields radiated into the cell from a horn antenna were measured using atomic field sensing methods. Measurement results were compared with simulations, and reasonable agreement was found. From the perspective of cell characterization, the main remaining issue is how exactly does the observed reduction of the DC electric field by about a factor of 3 occur. Future position-resolved spectroscopic field measurements may be helpful to address this issue.

Cells with wall-integrated electrodes present opportunities in science and technology. For instance,  
the DC quadrupole electric field investigated in the present work, as well as variations of it that employ all electrodes instead of only one electrode, combined with a longitudinal magnetic field, will enable studies of vapor-cell-based electron, ion and plasma Penning, cusp and Paul traps. Rydberg-EIT can be used to measure the electric fields in and around such charged-particle traps. HF and VHF radio-frequency signals directly applied via the wall-integrated electrodes may enable novel designs for Paul traps and mass spectrometers, and may become useful in advanced RF field sensing applications.      

\begin{acknowledgments}
This work was supported by NSF Grant Nos. PHY-1707377 and PHY-1806809 and Rydberg Technologies Inc. The cell was made in the Lurie Nanofabrication Facility at the University of Michigan. We thank Dr. Pilar Herrera-Fierro and her team for their technical support while developing the cell.
%We thank Dr. David Anderson for helpful discussions. 
%The cell has been filled with Rb and sealed at Rydberg Technologies Inc.  

\end{acknowledgments}

\appendix

\bibliography{penning}% Produces the bibliography via BibTeX.

\end{document}